\documentclass[twocolumn,showpacs,preprintnumbers,amsmath,amssymb]{revtex4}

\usepackage{graphicx}
\usepackage{dcolumn}
\usepackage{bm}

\begin{document}

\title{A new type of optical biosensor from DNA wrapped semiconductor graphene ribbons}

\author{Anh D. Phan}
\affiliation{Department of Physics, University of South Florida, Tampa, Florida 33620, USA}

\author{N. A. Viet}
\affiliation{Institute of Physics, 10 Daotan, Badinh, Hanoi, Vietnam}
\email{navietiop@gmail.com}

\date{\today} 

\begin{abstract}
Based on a model of the optical biosensors (Science 311, 508 (2006)) by wrapping a piece of double-stranded DNA around the surface of single-walled carbon nanotubes (SWCNT), we propose a new design model of this sensor, in which the SWCNT is replaced by a semiconductor graphene ribbon (SGR). Using a simple theory of exciton in SGRs, we investigated transition of DNA secondary structure from the native, right-handed B form to the alternate, left-handed Z form. This structural phase transition of DNA is the working principle of this optical biosensor at the sub cellular level from DNA and semiconductor graphene ribbons.
\end{abstract}

\pacs{Valid PACS appear here}
\maketitle

\section{Introduction}

Nowadays, the state-of-the-art achievements have been made at the frontier of nanotechnology and biotechnology by employing modern nanomaterials to manufacture biosensors \cite{1,2}. The paramount role of biosensors has covered a board range of clinical diagnosis, treatment method \cite{4}, and biomedical studies \cite{3}. Improving qualification of biosensors gives great challenges in terms of technology and requires the increase of understanding of biological world as well as new-found nanomaterials \cite{5}. 

The interactions between DNA and carbon systems has attracted a significant amount of attention from both experimental and theoretical scientists in order to develop sensitive biosensors. SWCNT and DNA were designed to study the B-Z change in living organisms \cite{6}, the relations between biomolecules and biosensors enhancing biostability \cite{7} and specificity \cite{8} of DNA. The graphene-DNA based biodevices as chemical sensor in order to detect disease was unveiled \cite{22}. Another research exploried that DNA leads to much less influence of enzymatic cleavage on graphene much due to the protection of DNA \cite{9}. Therefore, continuing to study the behaviors of graphene-DNA based biosensors and their applications is one of the hottest topics in this field. 
 
Graphene is a one-atom-thick planar sheet that has unique physical, mechanical and electronic properties. The exciton binding energies of the low dimensional systems have been studied over many years and can be clearly reported in the room temperature \cite{10,11,12}. This energy is the environment-dependent quantity. The measurement data of the exciton binding energy provides the information of environment. 

DNA is one of the most sophisticated and interesting objects in living organisms. There are many possible conformations of DNA. Until now, numerous reports have indicated that A-DNA, B-DNA, and Z-DNA have been found in nature \cite{13}. The DNA conformations are influenced by factors of the surrounding medium, for example, salt concentration \cite{6,13}. The configurations of DNA have the B-form at low salt concentration and the Z-form at the high salt concentration. Variation of environment changes the dielectric, size and other properties of DNA. This property gives the principle of behaviors of biosensors to detect the change of DNA structure as well as the environment change.

The simple model of biosensor in \cite{6} provides good predictions and agrees excellently with experimental data. In this letter, our modified biosensor consists of armchair graphene nanoribbon (AGNR) wrapped by DNA. Simple theoretical calculations and models of DNA are given in order to obtain the effective dielectric constant of medium. The link between the exciton energy transition and the dielectric constant of surrounding medium is presented. We proved that the AGNR-DNA based biosensors are more sensitive than CNT-DNA based biosensors in \cite{6}.

The rest of the paper is organized as follows: In Sec. II, the theoretical structure model of DNA is introduced in order to calculate the effective dielectric constant. In Sec. III, the exciton binding energy of GNR is presented. The principle of the GNR-DNA based biosensor and numerical results are shown in Sec. IV. Conclusions are given in Sec. V.

\section{Theoretical model of DNA}
\begin{figure}[htp]
\includegraphics[width=5.5cm]{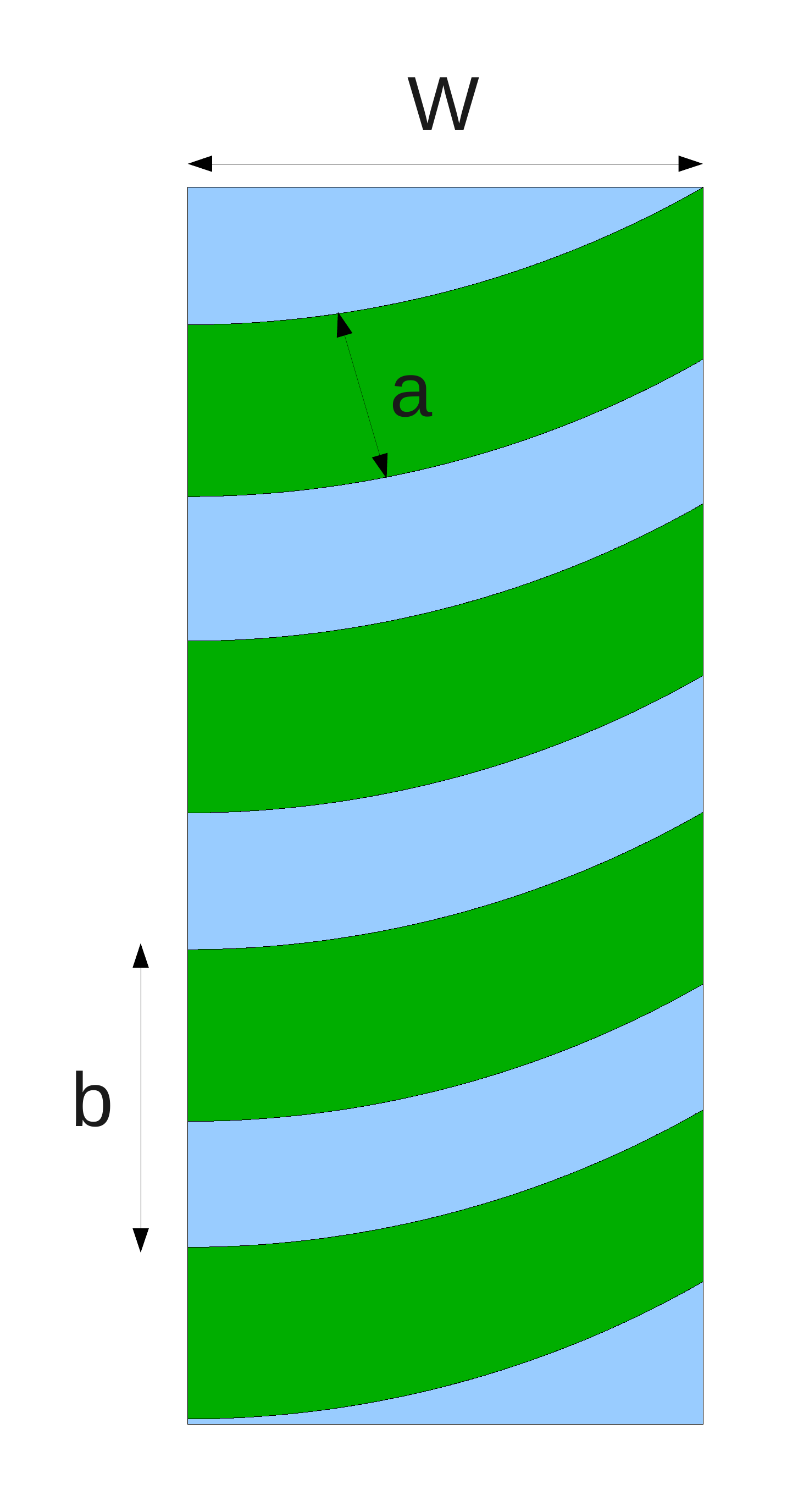}
\caption{\label{fig:1}(Color online) The scheme of AGNR-DNA based biosensor model.}
\end{figure}

In the present paper, we used a graphene nanoribbon (GNR) that has $W$ $nm$ in width and an atomic layer in thickness. The DNA strand is considered as a ribbon wrapping the GNR. The pitch along the axis of helical DNA is $b$, the width of DNA strand is $a$. Following previous work \cite{6}, the system is depicted in Fig.~\ref{fig:1}.

In this model, the effective dielectric constant of DNA and surrounding medium is given
\begin{eqnarray}
\varepsilon = f\varepsilon_{DNA} + (1-f)\varepsilon_{W},
\label{eq:1}
\end{eqnarray}
here $\varepsilon_{DNA}$ and $\varepsilon_{W}$ are the dielectric constants of DNA and solution, respectively, and $f$ is the ratio of DNA-covered surface area per total surface area of the GNR. Using the model of the model of helical ribbon for DNA, the expression of the ratio $f$ is obtained as a function of structural parameters of DNA and GNR
\begin{eqnarray}
f = \frac{al}{2bW},
\label{eq:2}
\end{eqnarray}
where $l$ is the straight length in a period of DNA. $l$ is approximately calculated by the following expression
\begin{eqnarray}
l = 2\sqrt{\frac{b^2}{4}+W^2}.
\label{eq:3}
\end{eqnarray}

For our calculation, it is assumed that DNA is a spring. It is easy to see that a position in $z$ axis is indicated
\begin{eqnarray}
z = \frac{b}{2\pi}\varphi.
\label{eq:4}
\end{eqnarray}

The total potential of DNA is generally presented by \cite{6}
\begin{eqnarray}
U_{tot} = U_{B} + U_{T} + U_{G},
\label{eq:5}
\end{eqnarray}
in which $U_{B}$ is the bending potential, $U_{T}$ is the torsional potential, and $U_{G}$ is the gravitational potential. $U_{G}$ can be ignored in nanoscale systems. The forms of $U_{B}$ and $U_{T}$ are performed
\begin{eqnarray}
U_{B} = \frac{1}{2}EI_{1}L(\kappa - \kappa_{0})^2, U_{T} = \frac{1}{2}GI_{2}L(\tau - \tau_{0})^2,
\label{eq:6}
\end{eqnarray}
here $E$ is the Young's modulus, $G$ is the shear modulus, $I_{1}$ and $I_{2}$ are the second moments of area of the spring, $L$ is the straight length of the spring, $\kappa_{0}$ and $\tau_{0}$ are the initial curvature and torsion of the spring, respectively. 

$\kappa$ and $\tau$ are calculated via other quantities
\begin{eqnarray}
\kappa = \frac{\varphi(L^2-z^2)^{1/2}}{L^2}, \tau = \frac{z\varphi}{L^2},
\label{eq:7}
\end{eqnarray}

Note that in a period, $\varphi = 2\pi$, $z = b$ and $L = l$, combining these parameters and Eq.(\ref{eq:3}), $\kappa$ and $\tau$ can be rewritten
\begin{eqnarray}
\kappa = \frac{4\pi W}{4W^2 + b^2}, \tau = \frac{2\pi b}{4W^2 + b^2},
\label{eq:8}
\end{eqnarray}
Substituting Eq.(\ref{eq:8}) into Eq.(\ref{eq:6}), one finds the total potential
\begin{eqnarray}
U_{tot} &=& \frac{1}{2}EI_{1}l\left(\frac{4\pi W}{4W^2 + b^2}-\frac{4\pi^2r_{0}}{4\pi^2r_{0}^2 + b_{0}^2}\right)^2 \nonumber \\
& & +\frac{1}{2}GI_{2}l\left(\frac{2\pi b}{4W^2 + b^2}-\frac{2\pi r_{0}}{4\pi^2r_{0}^2 + b_{0}^2}\right)^2,
\label{eq:9}
\end{eqnarray}
In the case of equilibrium system, minimizing the total potential energy of DNA yields
\begin{eqnarray}
b=\sqrt{\frac{W}{\pi r_{0}}(4\pi^2r_{0}^2+b_{0}^2)-4W^2},
\label{eq:10}
\end{eqnarray}
As a result, we finally obtain the expression of ratio $f$
\begin{eqnarray}
f=\frac{a}{2W}\sqrt{\frac{4\pi^2r_{0}^2+b_{0}^2}{4\pi^2r_{0}^2+b_{0}^2-4\pi r_{0}W}},
\label{eq:11}
\end{eqnarray}
here $r_{0}$, $b_{0}$ and $a$ are $1$, $3.32$ and $0.51$ nm, respectively, for B DNA; and $r_{0} = 0.9$ $nm$, $b_{0} = 4.56$ $nm$, and $a = 1.18$ $nm$ for Z DNA \cite{6}.
\section{Exciton energy of GNR}
We will consider a model of two-dimensional exciton formed by electron and hole. The interaction of electron-hole pair is known as the attractive Coulomb potential. The exciton Hamiltonian has the form \cite{11,12}
\begin{eqnarray}
H = -\frac{\hbar^2}{2m_{e}}\nabla_{e}^2 -\frac{\hbar^2}{2m_{h}}\nabla_{h}^2 + V(\vec{r_{e}}-\vec{r_{h}}), 
\label{eq:12}
\end{eqnarray}
here $m_{e}$, $\vec{r_{e}}$, $m_{h}$, $\vec{r_{h}}$ are the effective masses and positions of the electron and hole, respectively, $V(\vec{r_{e}}-\vec{r_{h}})$ represents the Coulomb interaction between the electron and hole. 

Introducing the coordinate of center of mass $\vec{R} = (m_{e}\vec{r_{e}}+m_{h}\vec{r_{h}})/(m_{e}+m_{h})$ and the relative $e-h$ position $\vec{r} = \vec{r_{e}}-\vec{r_{h}}$ for the motion in the x-y plane, the reduced mass of electron-hole pair $\mu = m_{e}m_{h}/(m_{e}+m_{h})$, the transformed Hamiltonian reads
\begin{eqnarray}
H = -\frac{\hbar^2}{2\mu}\nabla_{r}^2 -\frac{\hbar^2}{2(m_{e}+m_{h})}\nabla_{R}^2 + V(\vec{r}). 
\label{eq:13}
\end{eqnarray}
By using the separation of variables and solution of the Wannier model have been applied to the 2D semiconductor systems \cite{12} and AGNR systems \cite{19}, the exciton energy levels are give by
\begin{eqnarray}
E_{exc} = E_{g}-\frac{\mu e^4}{2\hbar^2 \varepsilon^2}\frac{1}{(n+1/2)^2} + \frac{\hbar^2 K^2}{2(m_{e}+m_{h})}, 
\label{eq:14}
\end{eqnarray}
where $E_{exc}$ is the exciton energy, $E_{g}$ is the band gap, $\varepsilon$ is the relative dielectric constant, $n$ is integers, and $\vec{K}$ is the wave vector that is obtained in the calculation due to the form of the envelope function $\psi(\vec{r},\vec{R}) = e^{i\vec{K}\vec{R}}\Phi(\vec{r})$. The second term is the binding exciton energy, denoted by $E_{B}$.  

Since the motion of the center of mass is extremely small, so we can approximate $\vec{K} = 0$ and $E_{exc} = E_{g} - E_{B}$. It is remarkable that the binding exciton energy is quite sensitive to the effect of the external fields. If the external interactions have larger energies than the exciton energy, the state of exciton can be broken. In terms of the 3D materials, $E_{B}$ is not large enough to observe in the room temperature. Therefore, one only reports the existence of the exciton effects in low temperatures. However, in the low dimensional systems, the exciton effect can be observed in $300$ $K$ because the quantum confinement creates an increase of overlap of wavefunctions. It means that the Coulomb interactions and the binding exciton energies are boosted. It can be easily realized that for the first exciton state of 2D materials ($ n= 0$), $E_{B}$ is four times larger than that of 3D materials. 

For GNR system, we have another approximation. The influence of edge effects on interactions is neglected. From this, the binding exciton energy of GNR can be calculated by the effective masses and the effective dielectric constant of medium. The larger dielectric constant is, the smaller $E_{B}$ is. 

It is essential to note that our GNR has a finite width but an infinite length. In the ref.~\cite{14}, authors pointed out the energy gap as a function of the number of carbon atoms across their width, namely, $N$. For $N = 5$, $N = 8$, $N = 11$, the energy gap is quite small, so GNRs can be metallic ribbons. For other value of $N$, the magnitudes of the energy gap are larger. Thus, these types of GNRs are able to be SGRs. An additional point is that the increase of the GNR width causes the decrease of energy gap.

Despite of non-stopping attempt of GNRs studies, manufacturing GNRs in the sub-10 nm regime is a challenging problem. The cutting-edge breakthoughs proved that one can produce AGNRs with ultrasmooth edges and the width below 10 nm \cite{15}, particularly the ultranarrow width of AGNRs down to 6 nm \cite{16}, 2-3 nm \cite{23,24}. In Ref. \cite{25}, authors used various chemical approaches to make the width of GNRs down to 1-2 nm. Therefore, it is possible to fabricate a type of sensor involving narrow width of AGNRs less than 3 nm. 
\section{Excitonic transition energy of biosensors in structural transformations in DNA}
\begin{figure}[htp]
\includegraphics[width=9cm]{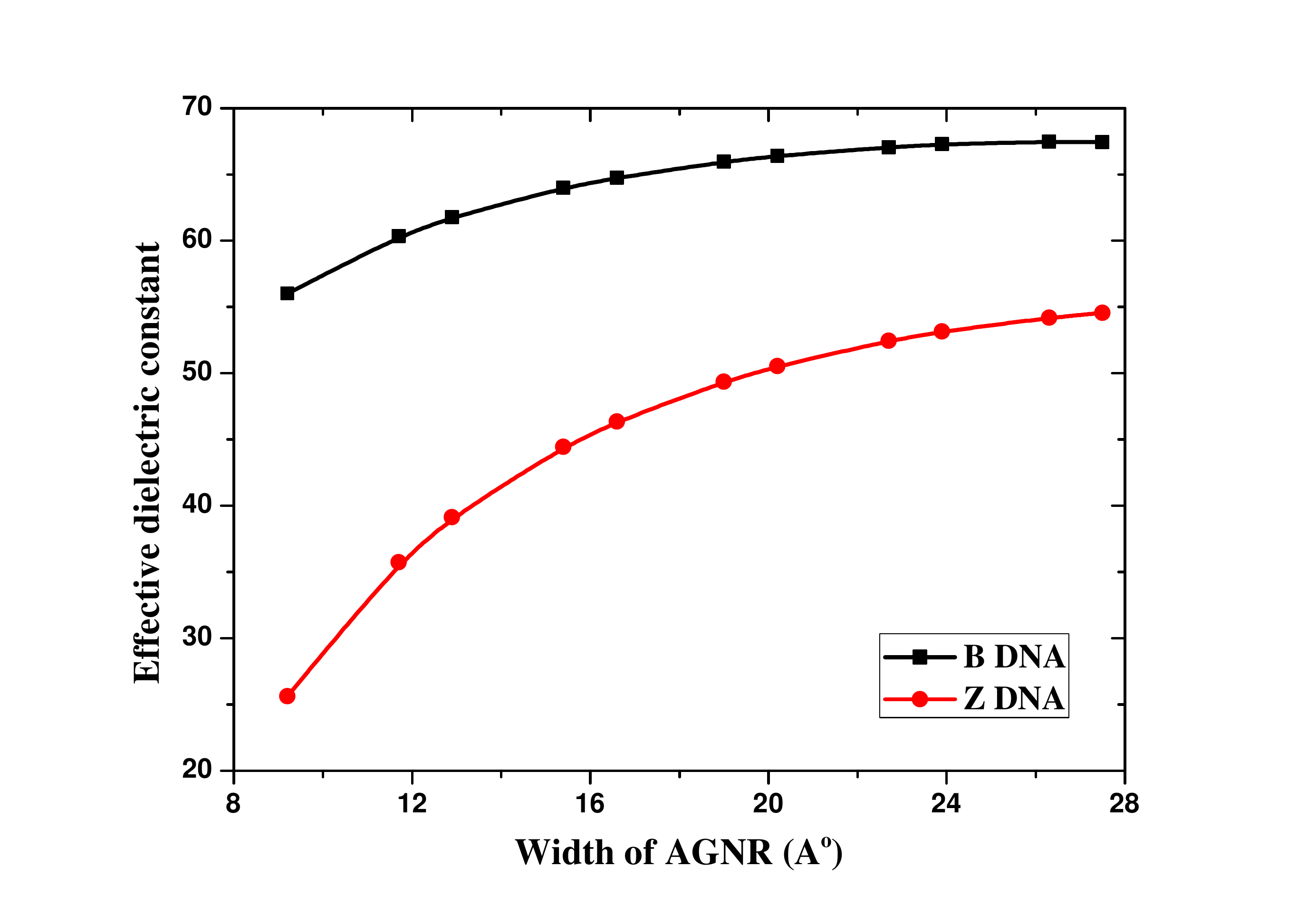}
\caption{\label{fig:2}(Color online) Effective dielectric constant of DNA and medium as a function of the AGNR width.}
\end{figure}

As mentioned above, DNA under the critical conditions has structural transformations. It can be explained by the interactions between negative surface charges of DNA and cations of solution. For different types of DNA, the critical concentrations of the same cation undergoing the B-Z structural transition have deviation \citep{13}. Another interesting point is that the higher valence cation is, the lower critical concentration is. When the structural transition occurs, the DNA area covering GNR changes. As a consequence, the ratio $f$ and the effective dielectric constant $\varepsilon$ vary. It leads to the change of the exciton binding energy. It is basis  in order to apply to design biosensor that can detect the relatively low concentration of high-valent metal caution below the cell scale of biology in blood or in biological cell. Fig.~\ref{fig:2} presents the width-dependent dielectric constants of DNA in the form of B and Z. In the calcualtion, the dielectric constant of DNA and water are $\varepsilon_{DNA} = 4.0$ \cite{6} and $\varepsilon_{W} = 80$ \cite{6,17}, respectively. 

It is important to note that when exciton effects and optical properties of the systems are calculated, the GNR should be semiconductor. The number of unit cells across the GNR width should be the form of $N = 3p$ or $3p + 1 $, here $p$ is an integer. Therefore, the width of GNR has to be discrete value instead of continous value. 

\begin{figure}[htp]
\includegraphics[width=9cm]{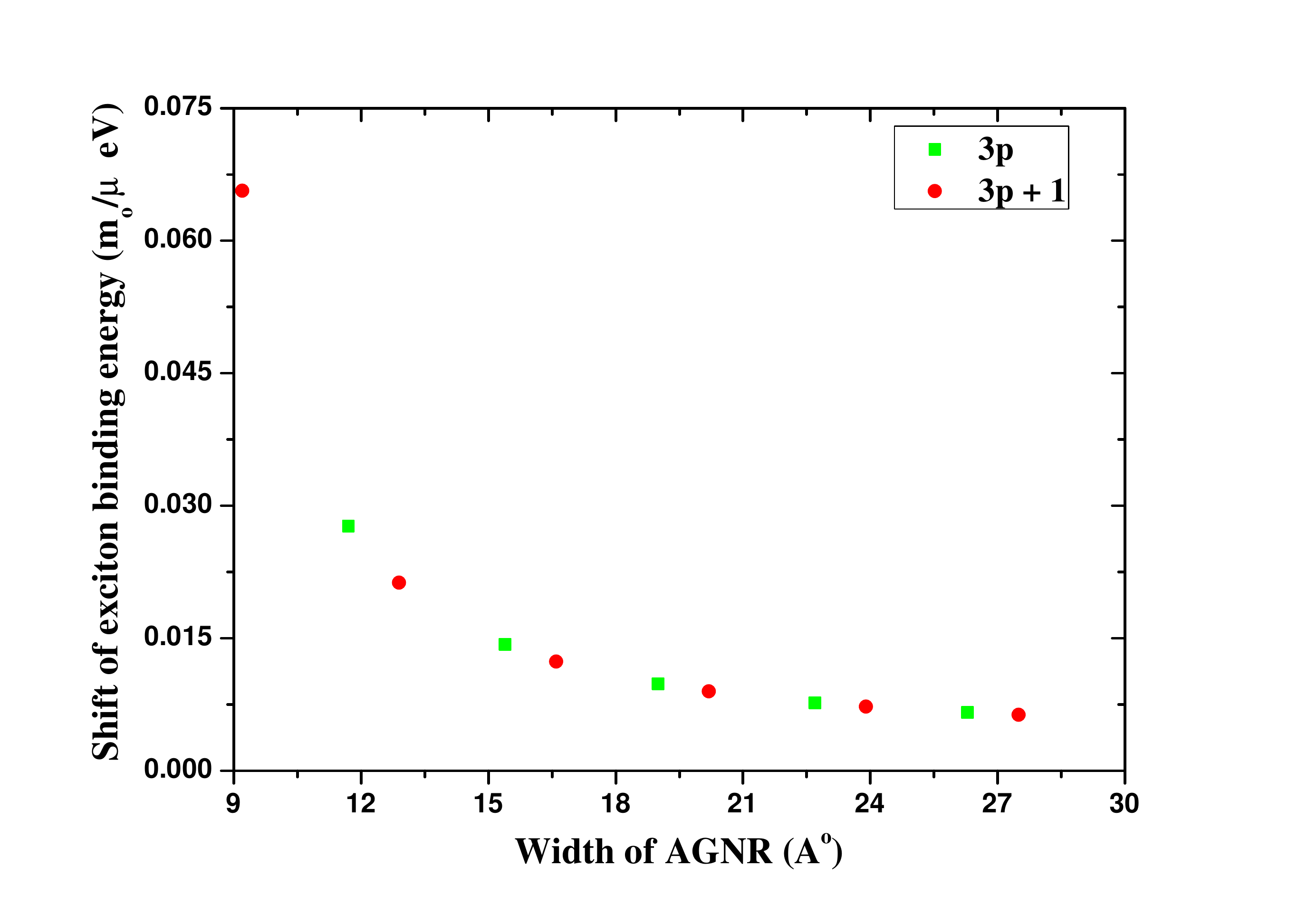}
\caption{\label{fig:3}(Color online) $\Delta E_{BZ}m_{0}/\mu$ in the B-Z structural transition as a function of the AGNR width.}
\end{figure}

Then, the exciton binding energy shift $\Delta E_{BZ}$ at $n = 0$ can be obtained 
\begin{eqnarray}
\Delta E_{BZ} = \frac{2\mu e^4}{\hbar^2}\left(\frac{1}{\varepsilon_{Z}^2}-\frac{1}{\varepsilon_{B}^2}\right) , 
\label{eq:15}
\end{eqnarray}
here $\varepsilon_{B}$ and $\varepsilon_{Z}$ are the effective dielectric constant when DNA is taken the form of B and Z, respectively. We have $\Delta E_{BZ}m_{0}/\mu$ in the B-Z structural transition in Fig.~\ref{fig:3}, where $m_{0}$ is the rest mass of electron. 

\begin{figure}[htp]
\includegraphics[width=9cm]{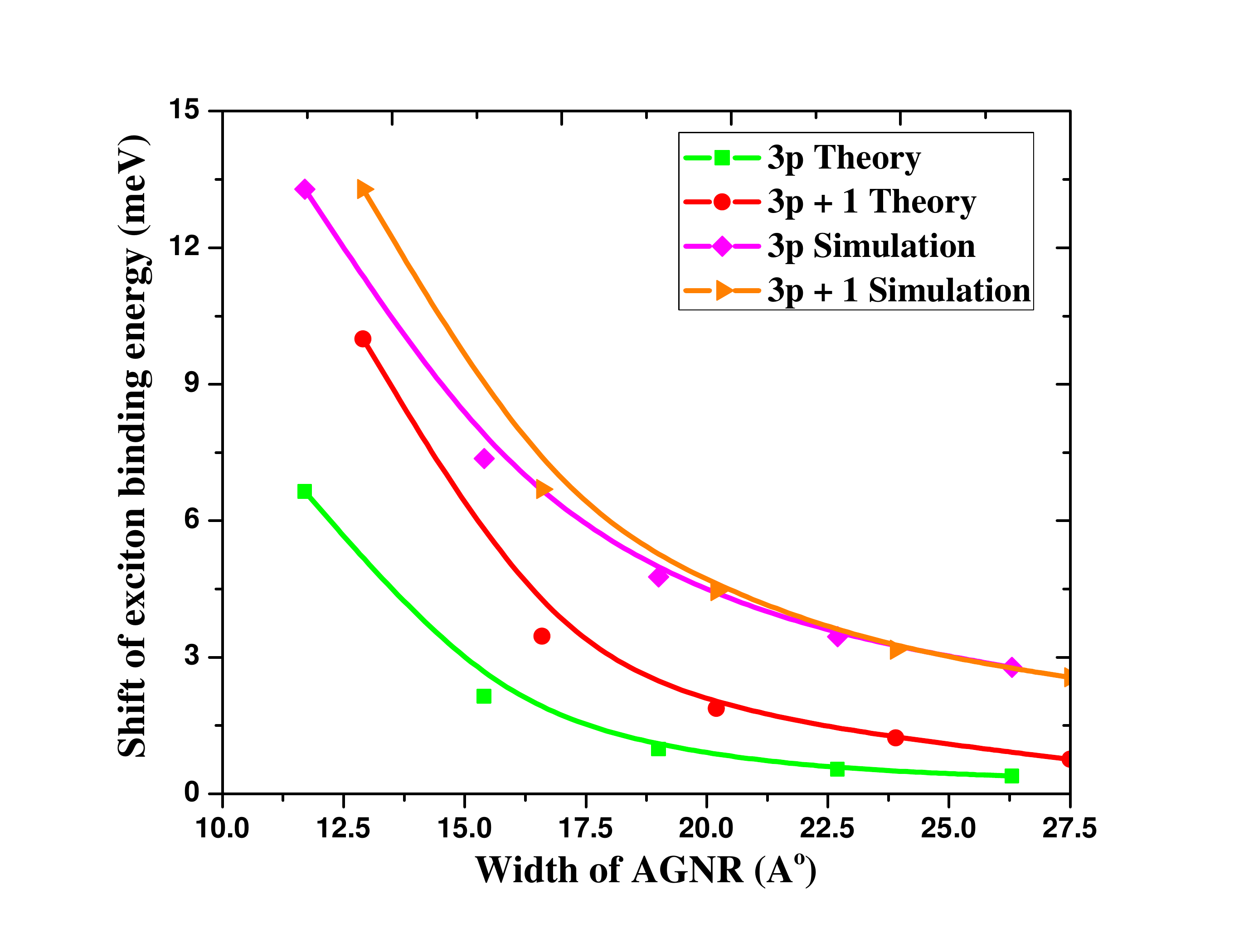}
\caption{\label{fig:4}(Color online) The shift of exciton binding energy in the B-Z structural transition as a function of the AGNR width.}
\end{figure}

In Fig.~\ref{fig:3}, $\Delta E_{BZ}m_{0}/\mu$ in the case of the number of the unit cells across the width $3p$ or $3p + 1$ is in the same curve. However, when we insert the effective mass of exciton of AGNR \cite{18} into the expression of $\Delta E_{BZ}$ and plot it in Fig.~\ref{fig:4}.

As we can see, the curves corresponding to $3p$ and $3p + 1$ are separated. A simple reason is that the effective mass of exciton depends on the AGNR width. Suppose that the periodic conditions are applied in AGNR, the band structure of AGNR is similar to the band structure of SWCNT with the diameter equal to the width of AGNR. Thus, the larger the width is, the smaller the effective mass is. The effective mass goes to zero when the width goes to infinity. The GNR infinite width means that GNR is graphene. The Ref. \cite{20} confirmed that the effective masses of electrons and holes of monolayer graphene are zero. The simple explaination has aggreement with the computational results in \cite{18} in terms of qualitative analysis.

The discrepancy between theoretical calculation and simulation calculation \cite{18} is due to the edge effect of AGNRs. As shown in Fig. \ref{fig:4}, when increasing the width of AGNRs, the curves of   in the case of 3p and 3p + 1 are closer and closer. The boundary conditions will disappear if  the AGNR width is large enough. A remarkable point is that   in the simulation calculations is larger than that of theoretical calculations. The exciton binding energy in simulation is approximately proportional to $\varepsilon^{-1}$  but Eq.4 points out that the energy is inverse proportional to  $\varepsilon^{2}$. It shows that the screening effect of electron in AGNRs is weak and plays an important role in the exciton binding energy. It also means that the exciton binding energy is much less dependent of environment \cite{18}.

In order to compare the exciton binding energy shift of AGNRs \cite{18} with that of CNTs, the values are chosen with the AGNR width nearly equal to CNT’s diameter. $\Delta E_{BZ}$ in the AGNR case is slightly smaller than that of CNT case.

GNRs can be considered as unzipped CNTs. It results that there are some similarities with the excitonic behavior in CNTs and GNRs. In the presence of magnetic field or electric field, the exciton binding energy and band gap of two carbon prototypes depend strongly on the fields \cite{21,26,27,28,29}. However, GNRs has characteristic properties, such as GNRs becomes a half-metallic material when in-plane external electric is applied \cite{28}, or there are spin polarization at the edge atoms of zigzag GNRs \cite{29}. The behaviors pave the way for unexpected applications and new generation devices. 

On the other hand, advanced techniques allow controlling effectively edge modifications of GNRs due to their planar structure. Some recent theoretical studies have investigated the optical properties of new type configuration of GNRs, so-called Chevron-type GNRs (CGNRs) \cite{30}. It is shown that the exciton binding energy of CGNRs is greater than that of regular GNRs. The result may lead to the increase of sensitivity of AGNRs based sensors.
\section{Conclusions} 
The combination between DNA and GNR is very interesting problem that sparks new studies in the future. By using a simple model for DNA, we have investigated the environment-dependent properties of biosensor. We showed the expression of effective dielectric constant of medium. In addition, we demonstrated that it is possible to record the B-Z structural transition via the change of the exciton binding energy. Our biosensor has good behavior when the width of AGNR is small.
\begin{acknowledgments}
The work was partly funded by the Nafosted Grant No. 103.06-2011.51.
\end{acknowledgments}

\newpage 

\end{document}